# Optical and Electronic Properties of Double Perovskite $Ba_2ScSbO_6$

Rajyavardhan Ray[1,], A K Himanshu[2*], Uday Kumar[3], Pintu Sen[2], J. Lahiri[2], and S K Bandyopadhyay[2], T. P. Sinha[4]

[1]Department of Physical Sciences, Indian Institute of Science Education and Research Mohali, Mohali-140306
[2] Variable Energy Cyclotron Centre, 1/AF, Bidhannagar, Saltlake, Kolkata-700 064, India
[3] Department of Physical Sciences, IISER Kolkata, Mohanpur Campus, Mohanpur, West Bengal, India
[4] Department of Physics, Bose Institute, 93/1 Acharya Prafulla Chandra Road, Kolkata-700 009, India

*akhimanshu@gmail.com, akh@vecc@gov.in  (Email of corresponding author)

**Abstract.** $Ba_2ScSbO_6$ (BSS) has been synthesized in polycrystalline form by solid state reaction at 1400 ˚C for 72 Hrs. Structural characterization of the compound was done through X-ray diffraction (XRD) followed by Riedvelt analysis of the XRD pattern. The crystal structure is cubic, space group Fm-3m (No. 225) with the lattice parameter, a = 8.230 Å. Optical band-gap of the present system has been calculated using the UV-Vis Spectroscopy and Kubelka-Munk (KM) function, it's value being 4.2eV. A detailed study of the electronic properties has also been carried out using the Full-Potential Linear Augmented Plane Wave (FPLAPW) as implemented in WIEN2k. BSS is found to be a large band-gap insulator with potential technological applications, such as dielectric resonators and filers in microwave applications.

## INTRODUCTION

Double Perovskite oxide of the general formula $A_2B'B''O_6$ is an example of strongly correlated electron system, and it has been the subject of investigation in solid-state and material science research in the recent years due to its broad class of properties such as high temperature superconductivity, colossal magnetoresistance, etc., have been studies since 1960s [1,2]. With the general formula $A_2BB'O_6$, the double perovskites are a three dimensional network of alternating octahedral $BO_6$ and $B'O_6$, with A-site atom occupying the interstitial space between these octahedral. Typically A-site atoms are alkaline earth metals or lanthanides and the B-site cations are transition metals. They constitute an important class of materials, characterized by structural distortions from the prototype cubic, space group Fm-3m (No. 225). These distortions are caused by rotations of the BO6 and B'O6 octahedra to accommodate the A-site cation.

Strontium Antimony double perovskite compounds have attracted considerable attention of late [3,4]. Recently, $Sr_2CrSbO_6$ (SCS) and $A_2ScSbO_6$ (A=Sr,Ca) have been synthesized and experimentally probed over a wide temperature range to study it's crystal structure and phase transitions [5,6]. The electronic structures of SCS, SSS and CSS have been studied before [7]. The ideal Double Perovskite has a cubic symmetry. This symmetry is usually expected when the tolerance factor t is close to unity. Here t is defined as $t=(r_A+r_O)/\sqrt{2}(r_{BB'} + r_O)$, where $r_A$, $r_{BB'}$ and $r_O$ are the atomic radii of the A-site cation, average of the B-site cations, and the O atom. For a majority of the double perovskite, however, the radius of the A-site cation is too small to fit the cavity formed by the oxygen octahedra and leads to a lower symmetry structure. The tolerance factor for the $Sr_2ScSbO_6$ (SSS) is 0.969, suggesting that the room temperature of these compounds should not be cubic. SSS is monoclinic having the space group P21/n at room temperature. With increasing temperature, SSS undergoes three structural phase transformations: P21/n → I2/m → I4/m → Fm-3m at approximately 400K, 550K and 650K.

In this article, we have performed the UV-vis spectroscopy and experimentally determined the optical band gap for these composition. We also initiated Density Functional Theory (DFT) calculations using Full Potential Linearized Augmented Plane Wave (FP-LAPW) using WIEN2k code for the first time for the composition Barium Scandium Antimony Oxide (BSS).

# EXPERIMENTAL AND COMPUTATIONAL DETAILS

Samples of BSS were synthesized using the solid-state reactions at 1400 ˚C for 72 hours. Color of the obtained sampled was found to be off-white. X-ray diffraction was carried out and was analyzed using Rietveld analysis software FULLPROF. The structure is found to be cubic, space group Fm-3m (No. 225) with lattice constant a=8.20Å. The positions of Ba, Sc, Sb, and O atoms are found to be *8c*, *4a*, *4b* and *24e* respectively with the x-coordinate for the O atom in *24e* position equal to 0.263.

In order to study the optical band-gap of the sample, diffusive reflectance measurement was carried out through UV-Vis spectroscopy (Perkin-Elmer 950). The Kubelka-Munk (KM) function was used to convert the reflectance spectra, thus obtained, into an equivalent absorption spectra [4,5]. The optical band-gap was obtained from the linear part of the energy dependence of the absorption spectra using an appropriate function, corresponding to direct-allowed transition. FIGURE 1 shows the absorption spectra and the corresponding value of band-gap. The value of the optical band gap was found to be approximately 4.2eV, as shown in the figure below (FIGURE 1.)

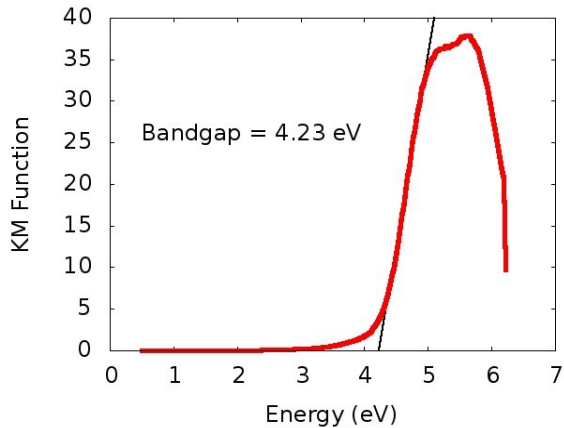

**FIGURE 1.** Linear dependence of the modified Kubelka-Munk (KM) function on the photon energy hυ corresponding to direct-allowed transitions, showing and optical band-gap of approximately 4.2eV.

Calculation of the electronic bandstructure for this complex double perovskite was performed using the Full Potential - Linear Augmented Plane Wave (FP-LAPW) method, in the framework of first principles Density Functional Theory (DFT), as implemented in WIEN2k. Starting with the experimentally obtained structure and atomic positions, volume optimization has been performed to obtain the fully-relaxed structure, obtained from the Birch-Murnaghan equation of state [5]. FIG. 2 shows the values of the energy corresponding to different values and well as the Birch-Murnaghan equation of state (dashed line). The values of optimal volume is found to be 934.06 a.u.$^3$. We use 72 k-points in the Brillouin zone and the muffin-tin radii for Ba, Sc, Sb, and O are, respectively, obtained to be 2.5, 2.22, 1.96 and 1.73. The density plane cut-off R*$k_{max}$ is 7.0, where $k_{max}$ is the plane-wave cut-off and R is the smallest of all atomic radii. The exchange and correlation effects have been treated within the Generalized Gradient Approximation (GGA). The self-consistency is better than 0.001 e/a.u.$^3$ for charge and spin density, and the stability is better than 0.01 mRy for the total energy per unit cell.

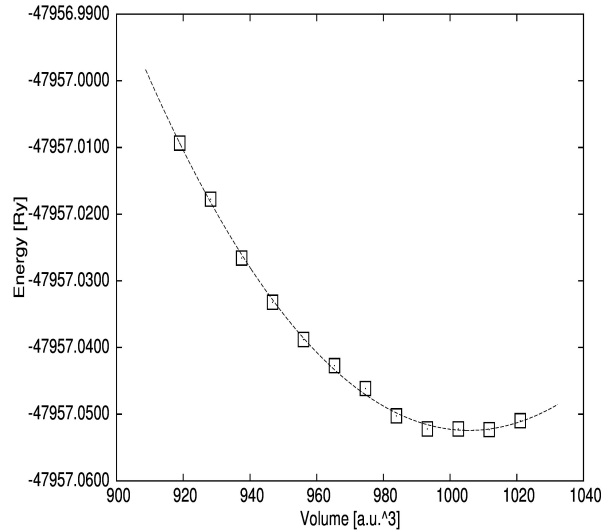

**FIGURE 2.** Energy versus volume for $Ba_2ScSbO_6$. The dashed curve is the Birch-Murnaghan equation of state.

We employ the Full-Potential Linear Augmented Plane Wave (FP-LAPW) method within the framework of density functional theory, as implemented in WIEN2k, to calculate the electronic structure. In this calculation, we have used the optimum values of the lattice parameters obtained from the experimental values [6,8]. The internal parameters of the atoms have been kept fixed at the experimental values. The optimum values of the lattice parameters are calculated from the ground state energies as a function of unit cell volume. The calculated results are then fitted using the Birch-Murnaghan equation of state [9]. FIG. 2 shows the values of the energy corresponding to different values and well as the Birch-Murnaghan equation of state (dashed line). The values of optimal volume is found to be 1005.6532 a.u.$^3$, corresponding to the lattice parameter 8.4160Å.

In this work, we study the room temperature cubic Fm-3m phase of BSS. We use 72 k-points in the Brillouin zone and the muffin-tin radii for Ba, Sc, Sb, and O are, respectively, obtained to be 2.5, 2.33, 1.75 and 1.75. The density plane cut-off R*$k_{max}$ is 7.0, where $k_{max}$ is the plane-wave cut-off and R is the smallest of all atomic radii.

Sc and Sb atoms are found to be in +3 and +5 valency respectively, leading respectively to $d^0$ and $d^{10}$ configurations. Thus, BSS is found to be a wide gap non-magnetic insulator.

## RESULTS AND DISCUSSIONS

FIGURE 3 shows the total and partial density of states, clearly exhibiting the insulating nature with a band of approximately 3.0eV. The discrepancy between the theoretical and experimental values of band gap is due to various approximations involved in the DFT calculations. As expected due to $d^0$-ness, the d-states of the Sc atoms lie in the conduction band. Similarly, the s and p-states of the Sb atom deep in the conduction band, as expected due to d-full-ness.

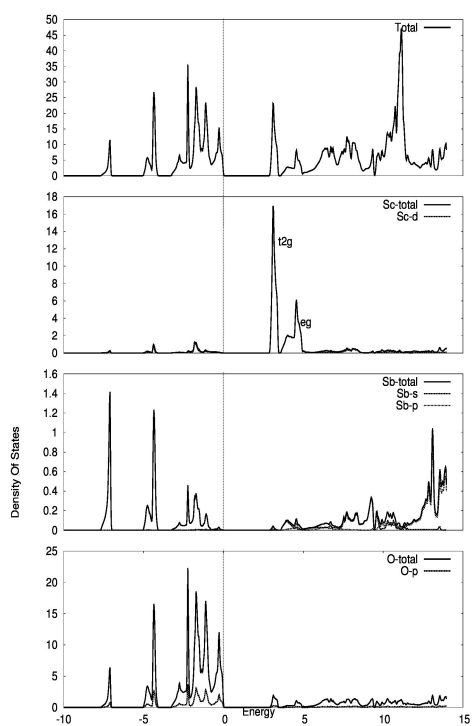

**FIGURE 3.** Total and partial spin-polarized density of states of Sc-d, Sb-s and p, and O-p in BSS.

The crystalline field produced by the oxygen octahedra lifts the degeneracy and leads to splitting of the d-states of Sc atom into $t_{2g}$ (consisting of dxy, dyz, dxz orbitals) and $e_g$ (consisting of $dx^2-y^2$ and $dz^2$ orbitals) states. Exchange splitting between the $t_{2g}$ and $e_g$ states is found to be approximately 1.43eV.

## CONCLUSIONS

In conclusion, we have experimentally obtained the optical band-gap and theoretically studied the electronic structure of BSS using FP-LAPW based first principles method. The ground state is found to be insulating with large band-gap.

Due to its large band-gap these material may be used as a dielectric resonators and filters in microwave applications. It would, therefore, be interesting to carry out the study of optical properties of such materials as optical applications such as interference filters, optical fibres and reflective coating require accurate knowledge of their optical response. At the same time, presence of $d^0$-ness and d-full-ness of the Sc and Sb atom makes it a potential functional material and a promising candidate for such.

## ACKNOWLEDGEMENTS

The authors AKH, JL, PS & SKB are thankful to Prof. D. K. Srivastava, Director of VECC for his kind interest and encouragement in the 12$^{th}$ year plan project (PIC No: XII-R&D-VEC-5.09.800).